\documentclass{elsart}

\usepackage{amsmath}
\usepackage{amssymb}
\usepackage{graphicx}

\begin{document}

\begin{frontmatter}



\title{Explicit predictability and dispersion scaling exponents in 
fully developed turbulence}


\author{Fran\c{c}ois G. Schmitt}

\address{CNRS, UMR 8013 ELICO, Wimereux Marine Station, University of Lille 1 \\
 28 av. Foch, 62930 Wimereux, France}

\begin{abstract}
We apply a simple method to provide explicit expressions for different scaling exponents 
in intermittent fully developed turbulence, that before were only given through a Legendre transform.
This includes predictability exponents for infinitesimal and non infinitesimal perturbations,
Lagrangian velocity exponents, and dispersion exponents. We obtain also new results concerning
inverse statistics corresponding to exit-time moments.
\end{abstract}

\end{frontmatter}

\section{Introduction}
 One of the main feature of homogeneous and isotropic fully developed turbulence,
 corresponding to very large Reynolds numbers, is the fact that statistical properties at
 small scales are scaling and universal \cite{Kolmogorov1941a}, and possess intermittent fluctuations
  \cite{Batchelor1949}. This is classically 
 characterized using structure functions scaling exponents,
 describing the probability density of velocity fluctuations at all scales belonging to the
 inertial range \cite{Frisch1995}.  
 This concerns the Eulerian velocity and its statistics, but the characteristic features of 
fully developed  turbulence is not only intermittency and multiple scales, it is also the loss of predictability,
 chaotic behaviour, dispersion and mixing properties, and a large number of degrees of 
 freedom. In the last ten years, many of these properties have been revisited 
  taking into account intermittency to consider how different scaling laws related to these issues
 are modified (for a review of several results in this framework see \cite{Boffetta2002b}). 
 The scaling exponents
 obtained have often been expressed as a Legendre transform of some function of
 velocity exponents, and numerical applications need a numerical Legendre transform involving
 the dimension function of velocity fluctuations. Here we present a simple method to simplify
 Legendre transform expressions and to replace them with parametric functions that provide
 explicitely the scaling exponents when the Eulerian velocity scaling exponent structure functions
 are known. In the following the general framework is presented in section \ref{sec2} and applied to
 different applications from predictability to $\epsilon$-entropy studies in section \ref{sec3}.
 Let us note that the results presented here correspond to the ideal situation of homogeneous and
 isotropic turbulence; for realistic flows, this property cannot be expected to hold perfectly since
 the influence of boundary conditions may introduce statistical inhomogeneity.

\section{The multifractal framework and Legendre transform}
 \label{sec2}
We recall first here the multifractal framework for velocity intermittency in turbulence, 
classically characterized using statistical moments of velocity increments. We recall the
Legendre transform introduced to relate scaling moment function to the codimension function of
velocity singularities. The latter is subsequently systematically used to simplify the Legendre transform
of some general expression of the codimension function. 

\subsection{Singularities, moment functions and Legendre transform}
In fully developed turbulence, intermittency is classically characterized by  $\zeta(q)$, the 
scaling exponent  of spatial structure functions. 
Denoting $\Delta V_{\ell} = \vert V(x+\ell)-V(x) \vert$ 
the increments of the velocity field at a spatial scale $\ell$, their 
fluctuations are characterized, in the inertial range, using the scale invariant 
moment function $\zeta(q)$ (see \cite{Frisch1995,Schertzer1997} for reviews):
\begin{equation}
<  \Delta V_{\ell}^q> \sim \ell^{\zeta(q)}
\label{eq1}
\end{equation}
where $q$ is the order of moment. Kolmogorov's initial proposal \cite{Kolmogorov1941a}, for a non-intermittent
constant dissipation, leads to $\zeta(q)=q/3$. For intermittent turbulence, $\zeta(q)$
is a cumulant generating function, and is nonlinear and concave; only the third moment has
no intermittency correction: $\zeta(3)=1$ (see below).
On the other hand, taking into account intermittency for the Eulerian velocity, one can write locally:
\begin{equation}
  \Delta V_{\ell}  \sim \ell^{h} 
  \label{eq2}
\end{equation}
where $h$ is a singularity of the velocity fluctuations. There is a continuous range of possible
$h$ values, and they are characterized through their codimension function 
$c(h)$ defined as \cite{Frisch1995,Schertzer1997}:
\begin{equation}
   p( \Delta V_{\ell} ) \sim \ell^{c(h)}
  \label{eq3}
\end{equation}
where $p( \Delta V_{\ell} )$ is the probability density of velocity fluctuations at scale 
$\ell$. Here the codimension is used for convenience, instead of the 
more classical dimension $f(h)=d-c(h)$,
where $d$ is the dimension of the space ($d=1$ or $d=3$ in most studies), and
$p( \Delta V_{\ell} )$ is the probability density of velocity fluctuations.  Let us note that,
while $h$ can be both negative or positive (but it is most of the time positive), the codimension function
$c(h)$ is positive and decreasing. This continuous range of $h$ values justified the use of
the term ``multifractal'' for such process \cite{Parisi1985}.

The moments write:
\begin{align}
   < \Delta V_{\ell} ^q> & = \int \Delta V_{\ell} ^q   p( \Delta V_{\ell} )    \\
                   & \sim  \int  {\ell} ^{qh+c(h)}   p(h) \sim \ell^{\zeta_E(q)} 
\end{align}
using a saddle point argument \cite{Parisi1985}, this gives the classical Legendre transform,
 between $\zeta_E(q)$ and 
the codimension function:
\begin{equation}
               \zeta(q) = \underset{h}{\min} \lbrace qh+c(h) \rbrace
      \label{eq4}
\end{equation}
This can also be written in the following way, emphasizing the one-to-one relation
between orders of moment $q$ and singularities $h$:
\begin{equation}
        \begin{cases}
               & qh+c(h)  = \zeta(q)     \\      
               & q  =-c'(h)    
          \end{cases}     
\label{eq5} 
\end{equation}
The last relation can be replaced by $h=\zeta'(q)$. Equations (\ref{eq5}) are used below.

\subsection{Estimates of structure function scaling exponents up to moments of order 8}
The scaling moment function $\zeta(q)$ is proportional to the cumulant generating function of the
variable $\log \vert \Delta V_{\ell} \vert $; it is nonlinear and concave. There are two ``fixed points'' for
this curve, corresponding to a piori exact values: $\zeta(0)=0$ and $\zeta(3)=1$. The
latter relation is a consequence of the Kolmogorov $4/5$ law \cite{Kolmogorov1941b}
for the third order longitudinal structure function (without absolute values):
\begin{equation}
   < \Delta V_{\ell} ^3>  =   - \frac{4}{5} \epsilon \ell
\label{eq6} 
\end{equation}
The other moments do not have well-accepted theoretical values; the analytical curves depend
on the stochastic model chosen for the statistics of velocity fluctuations. In agreement with
Kolmogorov's initial hypothesis concerning the universality of velocity fluctuations in the
inertial range, experimental estimates are rather stable for small and medium orders of
moments and do not depend on the Reynolds number: the exponents $\zeta_E(q)$ have been estimated experimentally for more
 than twenty years, for different types of flows (see 
 \cite{Anselmet1984,Benzi1993,Schmitt1994,Arneodo1996,Antonia1997,VdW1999}),
and are now considered as rather stable and almost universal until moments of order $7$ or $8$
\cite{Schmitt1994,Arneodo1996}. Let us note that the moment of order $7$ 
seems to be close to $2$, for any reasons that did not receive up to now any theoretical
founding.

Since these estimates are rather stable up to moments of order 8,
we chose in the following for simplicity for $\zeta_E(q)$ average values of the estimates published in 
Ref. \cite{Benzi1993,Arneodo1996,Antonia1997,VdW1999}: see Table \ref{tab1}.
These are estimates obtained from many experiments with $R_{\lambda}$ going from 35 to 5 000.
The experimental average which is used here is quite close to all these experimental estimates.

\section{Applications to different predictability and dispersion turbulence functions}
 \label{sec3}

\subsection{Predictability of small perturbations in the intermediate dissipation range of turbulence}
Let us first consider the predictability property of small perturbations in turbulence.
It is well-known that the viscous cutoff, corresponding to the scale for which the
Reynolds number is of order 1, is different from the Kolmogorov scale $\eta$ when
intermittency is taken into account: more intense singularities have
a smaller local cutoff scale and weak fluctuations have larger cutoff scales. This
range of scales is the intermediate dissipation range, where there is a mixture of inertial
and dissipation effects.
Using Eq. (\ref{eq2}) and the Reynolds number
$Re=VL/\nu$, this writes \cite{Paladin1987}:
\begin{equation}
   \eta_h=L Re^{-1/(1+h)}
\label{eq7} 
\end{equation}
where $L$ is a large scale and $\eta_h$ is the cutoff scale associated to singularity $h$. 
This spatial scale is
associated to a temporal cutoff scale $\tau_h$ through the velocity $v_h=(\eta_h)^h$:
$\tau_h=\eta_h/v_h=(\eta_h)^{1-h}$ giving the singularity dependent cutoff time scale:
\begin{equation}
   \tau_h=T Re^{\frac{h-1}{h+1}}
\label{eq8} 
\end{equation}
where $T$ is the large time scale associated to $L$. This temporal cutoff scale is the smallest 
scale of the system: below these scales viscosity dominates. In 
fully developed turbulence, the positive maximum Lyapunov exponent
$\lambda$, characterizing the exponential growth rate of an infinitesimal disturbance, is
proportional to the inverse of the smallest characteristic time of the system. Since there is 
a range of time scales associated to
the continuous range of $h$ values through Eq.(\ref{eq8}), the Lyapunov exponent may be given by
\cite{Crisanti1993a,Crisanti1993b}:
\begin{equation}
   \lambda \sim \int \frac{1}{\tau_h}dp(h) \sim \int Re^{\frac{1-h}{h+1}}Re^{\frac{-c(h)}{h+1}}dh
\label{eq9} 
\end{equation}
giving finally:
\begin{equation}
   \lambda \sim Re^{\alpha}
\label{eq10} 
\end{equation}
with the exponent $\alpha$ given by a saddle-point argument since the Reynolds number is large \cite{Crisanti1993a,Crisanti1993b}:
\begin{equation}
               \alpha = \underset{h}{\max} \left\{ \frac{1-h-c(h)}{1+h}\right\}
      \label{eq11}
\end{equation}
This was solved numerically by Crisanti et al. \cite{Crisanti1993a} to provide 
$\alpha \simeq 0.46$
through a numerical Legendre transform using experimental estimates of the function $c(h)$.
This result is slightly smaller than the one provided by Ruelle \cite{Ruelle1979} for
nonintermittent turbulence: $\alpha=1/2$.

Let us go further and show here how to obtain this exponent, considering, for more generality, the
moments of order $q>0$ of the inverse times:
\begin{equation}
               < \frac{1}{\tau_h^q}  >  \sim Re^{\alpha(q)}
      \label{eq12}
\end{equation}
where, using as done above a saddle-point argument, the exponents $\alpha(q)$ are given by:
\begin{equation}
               \alpha(q) = \underset{h}{\max} \left\{ \frac{q(1-h)-c(h)}{1+h}\right\}
      \label{eq13}
\end{equation}
 In general, the singularity $h_0$ for which this 
condition is met verifies $G'(h_0)=0$ where 
\begin{equation}
               G(h) = \frac{q(1-h)-c(h)}{1+h}
                     \label{eq14}
\end{equation}
Estimating the derivative $G'(h_0)=0$ and using Eq. (\ref{eq5}) to introduce the moment $q_0$ associated to the singularity $h_0$ leads to:
\begin{equation}
               \zeta(q_0)=2q-q_0
   \label{eq15}
\end{equation}
This provides a unique value of $q_0$  and hence of $h_0$, for a given value of $q$. This equation,
together with another use of the Legendre transform, gives the
function $\alpha(q)=G(h_0)$:
\begin{align}
    \alpha(q) = G(h_0) & = \frac{q(1-h_0) -(\zeta(q_0)-q_0h_0)}{1+h_0}   \notag \\
                                       & = q_0-q \notag 
\end{align}
This gives finally the following parametric relation between $\alpha(q)$ and the moment function
$\zeta(q)$:
\begin{equation}
        \begin{cases}
               & \zeta(q_0)=2q-q_0      \\      
               &  \alpha(q)  = q_0-q
           \end{cases}     
\label{eq16} 
\end{equation}
The nonintermittent curve obtained for $\zeta(q)=q/3$ is $\alpha(q)=q/2$.
The explicit parametric relation (\ref{eq16}) can be used
to represent the curve $\alpha(q)$ for experimental estimates of $\zeta(q)$: this is shown in
Fig. 1 using the average $\zeta(q)$ curve. One can see from Fig. 1 that up to moments of order $2.5$ the 
linear non-intermittent curve is a very good approximation. The value of $\alpha(1)=\alpha$
is smaller than $1/2$ (close to $0.48$) but the intermittent correction is very small.
For $q=1$ we have $\alpha=\alpha(1)=q_0-1$, where $q_0$ is given by $\zeta(q_0)=2-q_0$.
For $q=2$ one obtains $q_0=3$ since $\zeta(3)=1$, and hence $\alpha(2)=1$ is
a fixed point, non affected by intermittency, recovering a result
already given in Ref. \cite{Crisanti1993a}. One can also obtain the following estimates:
$\alpha(3) \simeq 1.56$, $\alpha(4) \simeq 2.2$, $\alpha(4.5) \simeq 2.5$ 
(coming from $\zeta(7) \simeq 2$), and $\alpha(5) \simeq 2.85$.

\subsection{Predictability of noninfinitesimal perturbations in turbulence}
The same approach can be used to express scaling exponents characterizing
time scale statistics associated to noninfinitesimal perturbations.
As proposed in Ref. \cite{Aurell1996} and developed in Ref. \cite{Boffetta2002b}, and using the
notations of the latter reference, we consider a perturbation of size $\delta$ of the
velocity field, for $\delta$ belonging to
 the inertial range (hence the term noninfinitesimal). The time scale associated
to the eddy of typical velocity $\delta$ can also be considered as the decorrelation time
associated to the perturbation. In this framework, the inverse time is $1/\tau \sim \delta/\ell$
and since using the singularity $h$, one has $\ell \sim \delta^{1/h}$, we obtain locally
the relation between the time scale and the perturbation $\delta$:
\begin{equation}
                \frac{1}{\tau}    \sim \delta^{\frac{h-1}{h}}
      \label{eq17}
\end{equation}
As before, this can be used to consider the moments of order $q>0$ of the inverse times,
but here we will consider their scaling property as a power-law function of the velocity
perturbation $\delta$. Using the same type of argument as above, the moments of the inverse time
are expressed as an integral over all singularities $h$ an we consider all functions as
power-law functions of $\delta$ so that using a saddle-point argument, one has a scaling law:
\begin{equation}
               < \frac{1}{\tau^q}  >  \sim \delta^{-\beta(q)}
      \label{eq18}
\end{equation}
where  the exponents $\beta(q)$ are given by the following relation (using the fact
that the probability density writes $p(h) \sim \ell^{c(h)} \sim \delta^{c(h)/h}$):
\begin{equation}
              - \beta(q) = \underset{h}{\max} \left\{ \frac{q(h-1)+c(h)}{h}\right\}
      \label{eq19}
\end{equation}
where the negative sign has been introduced for convenience to have positive final numerical values
for $\beta(q)$ (see below). As done above for infinitesimal perturbations, this can be expressed
explicitely introducing the singularity $h_0$ that maximises the expression into brackets. 
With the same procedure as above: involving a differenciation and the introduction of the moment
of order $q_0$ associated to $h_0$, we obtain the final result as a parametric relation
between $\beta(q)$ and $\zeta(q)$:
\begin{equation}
        \begin{cases}
               &  \zeta(q_0)  = q        \\      
               &  \beta(q)=q_0-q =\zeta^{(-1)}(q)-q
           \end{cases}     
\label{eq20} 
\end{equation}
The exponent $\beta(q)$ is simply linked to the reciprocal of $\zeta$, denoted
$\zeta^{(-1)}$.
The nonintermittent curve obtained for $\zeta(q)=q/3$ is $\beta(q)=2q$.
Using Eq.(\ref{eq20}) we can see that $\zeta(3)=1$ gives $\beta(1)=2$, which
is a fixed point for $\beta(q)$, non affected by intermittency corrections \cite{Aurell1996}.
Figure 2 shows the resulting function compared to the straight line of equation
$2q$. It is visible that $\beta(q)$ grows very fast. Indeed, 
the relation $\zeta(7) \simeq 2$ gives the value $\beta(2) \simeq 5$. 
We can see that for $q<1$, $\beta(q)<2q$ and for $q>1$, $\beta(q)>2q$.
The result
$\beta(1)=2$ was checked using shell-model simulations \cite{Boffetta2002b}.

\subsection{Lagrangian velocity structure functions' in turbulence}
As an analogy with Kolmogorov's dimensional analysis in the Eulerian framework, Landau
\cite{Landau1944} proposed  in 1944  a $1/2$ law for the temporal increments
of the Lagrangian velocity $\Delta V_{\tau} = \vert V(t+\tau)-V(t) \vert$. This was later generalized
by Novikov, with a Lagrangian intermittency framework for the velocity \cite{Novikov1989}:
\begin{equation}
<  \Delta V_{\tau}^q> \sim \tau^{\zeta_L(q)}
\label{eq21}
\end{equation}
As for the Eulerian case, for a constant dissipation one obtains the ``mean field''
expression, neglecting intermittency: $\zeta_L(q)=q/2$. In this framework, the third order
moment for the Eulerian velocity is analogous to the second order moment for the
Lagrangian velocity: in case of intermittency $\zeta_L(q)$ is nonlinear and concave, and
the non-intermittent function is valid only for $q=2$: $\zeta_L(2)=1$.

Recently some authors have proposed an hypothesis helping to relate Eulerian and Lagrangian
structure functions \cite{Boffetta2002a,Biferale2004,Biferale2005,Chevillard2003}, following an earlier proposal by Borgas \cite{Borgas1993}. 
They consider the velocity advecting Lagrangian trajectories as a superposition of 
different velocity contributions coming from different eddies having different characteristic times.
After a time $\tau$ the fastest eddies, of scale smaller than $\ell$, are decorrelated so that at
leading order, they assume \cite{Boffetta2002a,Biferale2004,Biferale2005}:
\begin{equation}
\Delta V_{\tau} \sim  \Delta U_{\ell}
\label{eq22}
\end{equation}
writing $ \Delta U_{\ell} \sim \ell/\tau$ and using $\Delta U_{\ell} \sim \ell^h$, gives
the time and space local correspondence $\tau \sim \ell^{1-h}$.
Introducing this relation inside the integral of Eq.(\ref{eq5}),  
one obtains for the moments of the Eulerian velocity increments:
\begin{equation}
               < \Delta U_{\ell} ^q>  \sim  \int  \tau^{ \frac{qh +c(h)}{1-h}} dp(h)
 \label{eq23}
\end{equation}
Using a saddle point argument and using Eqs.(\ref{eq22}) and (\ref{eq21}), this gives the Lagrangian 
structure function as a Legendre transform \cite{Boffetta2002a}:
\begin{equation}
                \zeta_L(q) = \underset{h}{\min} \left( \frac{qh+c(h)}{1-h} \right) 
                \label{eq24}
\end{equation}
The authors that explored this relation did not go further to express $\zeta_L(q)$ in the general case \cite{Boffetta2002a,Biferale2004,Biferale2005}.

In fact, as we showed elsewhere for a Lagrangian turbulence
study \cite{Schmitt2005}, using an approach analogous to the ones above, one can obtain an explicit relation
between Eulerian and Lagrangian structure functions' scaling exponents.
We introduce the singularity $h_0$ that maximises the expression into brackets,
and with the same procedure involving a differenciation and the introduction of the moment
of order $q_0$ associated to $h_0$ we obtain the final result (see also \cite{Schmitt2005}):
\begin{equation}
        \begin{cases}
               & \zeta_L(q)  = \zeta_E(q_0)    \\      
               & q  = q_0-\zeta_E(q_0)
           \end{cases}     
\label{eq25} 
\end{equation}
This relation provides $\zeta_L(q)$ when the fonction $\zeta_E(q)$ is known, the second
 lign giving the link between $q$ and $q_0$. We can check that for $q_0=3$, the second
 line gives $q=2$ and the first one, $ \zeta_L(2)  = \zeta_E(3) =1$. We can also consider the
 case $q_0=7$ and using $\zeta_E(7) \simeq 2$, the approximate result $\zeta_L(5) \simeq 1$.

  Let us note that an analogous expression is provided in \cite{Borgas1993} for singularities and 
 (with our notations) codimension functions $c(h)$ in his Appendix A (Equation (A 4)). 
 Borgas used for this other types of arguments, including a Reynolds number scaling expression 
 for Eulerian and Lagragian statistics; however, he did not go to the moment framework
 to provide an equivalent expression for structure functions scaling exponents.
 Let us also note that some comparisons are performed in \cite{Boffetta2002a}
between their shell model Lagrangian values and
Eulerian shell model estimates, transformed through a numerical Legendre transform using the
analytical expression of a fit of $c(h)$.
The same type of comparison is performed in
\cite{Biferale2005} between their DNS estimates and a numerical Legendre transform of
Eulerian values. The approach proposed here is more direct since we test an explicit
 parametric relation corresponding to different hypothesis.

 The Euler-Lagrange relation given here is non-linear, so that one cannot
 expect to obtain the same type of statistical models in Eulerian and Lagrangian frameworks.
 For example, as explained by Chevillard \cite{Chevillard2003} using Borgas' relation, in this
 framework, the statistics cannot be lognormal for both Eulerian and Lagrangian velocity fields.
 Indeed, for $\zeta_E(q)=aq-bq^2$ with $a=(2+\mu)/6$ and $b=\mu/18$, one obtains
 $\zeta_L(q)=(a-1+\sqrt{(1-a)^2+4qb}-2bq)/2b$. For more details see our paper
 \cite{Schmitt2005} where different proposals for the Lagrangian structure functions scaling
 exponents are tested.
 
 In Figure 3, the Lagrangian prediciton is compared to Lagrangian estimates published in recent papers 
\cite{Boffetta2002a,Mordant2004,Biferale2004}. These papers report different types of such estimates:
Ref. \cite{Mordant2004} gives the values of experimental estimates for four flows going from
$R_{\lambda}$ of 510 to 1 000, and Direct Numerical Simulations (DNS) estimates 
for $R_{\lambda}$=75 and 140. Ref. \cite{Boffetta2002a} reports shell model estimates and
Ref. \cite{Biferale2005} DNS corresponding to $R_{\lambda}$=280. In order not to have a too heavy
figure, we plotted in Figure 3 four different types of Lagrangian values: 
an average value of the 4 experiments shown
in Ref. \cite{Mordant2004}, the DNS simulation corresponding to $R_{\lambda}$=140 of the same
publication, the shell model estimates of Ref. \cite{Boffetta2002a} and the DNS estimate of 
Ref. \cite{Biferale2005}. 
The average Eulerian curve is transformed and compared to these four 
Lagragian estimates. Several comments can be proposed from Figure 3. 
First, let us note that these experimental
and numerical estimations for Eulerian and Lagrangian scaling exponents come from different
types of flows with different Reynolds numbers. For Eulerian estimates, the exponents have been
carefully compared and are now considered as rather stables up to moments of order 8, but
for Lagrangian estimates, this work is still to do; measurements are only available for a few years,
and the scaling ranges are still small, so that the range of variability for these exponents is 
expected to narrow in the future. Indeed, as for the Eulerian case, one can expect that for high
Reynolds numbers these exponents are rather stable and universal, up to a given order of moment
that may be smaller than the Eulerian case, for which this critical order of moments
is about 7-8.  

With these comments in mind, we can provide some preliminary
conclusions from the observation of Figure 3: (i) up to moment of order 3, the 
prediction are close to all Lagrangian data; (ii) for larger moments, there is too much scatter in the
estimates to conclude about the validity of the transformation given by Eq.(\ref{eq25}).
A possible explanation of such scatter is the fact that scaling can be strongly affected by trapping
in vortex filaments \cite{Biferale2005}. In this case, after a filtering of high frequencies, the data
may be closer to the multifractal prediction.

\subsection{Relative turbulent dispersion}
We next consider the influence of intermittency on relative dispersion: this deals with the
statistics of particle pair dispersion, advected by an intermittent Eulerian velocity field.
This topic is close to the Lagrangian case. let us consider $R(t)$, the distance between a pair
of particles advected by an intermittent turbulence. Following Ref. 
\cite{Novikov1989,Boffetta1999a,Boffetta1999b}, one can introduce $\Delta V_{\tau} (R)$,
the velocity difference along the Lagrangian trajectories. This can be written:
\begin{equation}
               \frac{dR}{dt} =  \Delta V_{R} 
 \label{eq26}
\end{equation}
where $ \Delta V_{R} $ is the increment of the Lagrangian 
velocity is evaluated for a distance $R$ associated
to a time scale $\tau = R/\Delta V_{R} $. 
Then using the same hypothesis as for the Lagrangian case (Eqs. (\ref{eq21}) and (\ref{eq22})),
we obtain the time evolution of the moments of the pair distance:
\begin{equation}
              < R^q>  \sim < \left( \tau  \Delta V_{\tau} \right)^q > \sim \tau^{H(q)}
 \label{eq27}
\end{equation}
where $H(q)$ is the scaling exponent characterizing the intermittent pair dispersion. 
Using Eq.(\ref{eq24}) the exponents $H(q)$ can be given as a Legendre transform
of the Eulerian velocity, as provided by Boffetta et al. \cite{Boffetta1999a,Boffetta1999b}:
\begin{equation}
                H(q) = \underset{h}{\min} \left( \frac{q+c(h)}{1-h} \right) 
                \label{eq28}
\end{equation}
In fact Eq.(\ref{eq27}) shows also directly that the dispersion exponent is simply
related to the Lagrangian velocity exponent:
\begin{equation}
                H(q) = q+\zeta_L(q)                
    \label{eq29}
\end{equation}
Equations (\ref{eq25}) and (\ref{eq29}) then give:
\begin{equation}
        \begin{cases}
               & H(q)  = q_0  \\      
               & q  = q_0-\zeta_E(q_0)
           \end{cases}     
\label{eq30} 
\end{equation}
This can also be written $H(q-\zeta(q))=q$.
This relation could also have been obtained from Eq.(\ref{eq28}) using the same approach as above.
With $q_0=3$ we obtain $q=2$ and $H(2)=3$, corresponding to Richardson's law 
$<R^2> \sim \tau^3$ \cite{Richardson1926},
which is thus a fixed point for which there is no intermittency correction, as already noticed
\cite{Novikov1989,Boffetta1999a,Boffetta1999b}.
In Fig. 4, we represent the transform (\ref{eq30}) compared to the non-intermittent
line of equation $3q/2$, compared also to the shell model estimates published
in Ref. \cite{Boffetta1999b}. This shows that the nonlinearity of $H(q)$ does not
seem very strong. Furthermore, $H$ is concave, and the values give by
Eq. (\ref{eq30}) are quite close to the ones reported in Ref. \cite{Boffetta1999b}.

\subsection{Exit-time moments and $\epsilon$-entropy}
We finally consider exit-time moments \cite{Biferale1999},  also called inverse structure
functions \cite{Jensen1999}. We consider a time series of a turbulent quantity
$X(t)$ having intermittent fluctuations characterized by singularities $h$ such as
$\Delta X_{\tau} \sim \tau^h$ with codimension $c(h)$:
$p(h) \sim \tau^{c(h)}$. The moments are assumed to scale as 
$< \Delta X_{\tau}^q > \sim \tau^{\zeta(q)}$ with Eq.(\ref{eq7}). One then studies the dynamics
of such signal by considering exit times (also called distance structure functions
or inverse turbulence functions): considering a threshold value
$\delta = \Delta X_{\tau}$, let us denote $\tau(\delta)$ the first time needed
to obtain an increment $\delta = \Delta X_{\tau}$. A new time series
$(\tau_i)$ can be obtained this way, whose statistics will scale as the threshold value $\delta$:
\begin{equation}
        < \tau^q> \sim \delta^{\chi(q)}
     \label{eq31} 
\end{equation}
where $\chi(q)$ can be obtained, as before, as a Legendre transform involving $c(h)$. Writing
$\tau \sim \delta^{1/h}$ and inserting this into the integral of the moment
estimation gives \cite{Jensen1999,Biferale1999}:
\begin{equation}
                \chi(q) = \underset{h}{\min} \left( \frac{q+c(h)}{h} \right) 
      \label{eq32}
\end{equation}
Here as before, we estimate this function explicitely. A derivative of $G(h)=(q+c(h))/h$ leads
to $\chi(q)=G(h_0)$ with $\zeta(q_0)=-q$. 
We then obtain 
\begin{equation}
        \begin{cases}
               & \chi(q)  = -q_0  \\      
               & \zeta(q_0) = -q
           \end{cases}     
\label{eq32b} 
\end{equation}
The result then writes simply:
\begin{equation}
       \chi( -\zeta(q_0)) = -q_0
     \label{eq33} 
\end{equation}
which is exactly a result given in Ref. \cite{Roux2004} based on an exact result for a special
case (a multifractal Cantor set), and experimentally verified using shell model turbulence simulations, but
with no formal proof of the formulae in the general case. A similar expression obtained in the very different
context of Laplacian random walk is given in Ref. \cite{Hastings2002}. 

We must note here,
as already noticed in \cite{Roux2004}, that Eq.(\ref{eq33})  involves for either the original series
or for the return time series, negative moments. 
Let us consider first positive moments of return times: $q>0$ in Eq.(\ref{eq31}).  
We need here the hypothesis of the existence of negative moments of the
velocity structure functions, but it may not be the case, since this needs in practice
extremely precise measurements.  It is argued in \cite{Pearson2005} that negative moments for
$q<-1$ do not exist since the pdf of velocity increments does not vanish at $0$. 
The derivative writes $G'(h_0)=-(q+\zeta(q_0))/h_0^2$. For $q>0$, if negative moments $q_0$
such that $\zeta(q_0)=-q<0$ are not accessible in the experimental time series, then the 
derivative stays strictly negative, $G(h)$ does not reach a relative minimum, 
and the minimal value selected
by Eq.(\ref{eq32}) will be obtained for the maximal value of $h$, which is 
a ``minimal'' singularity of the velocity fluctuations, and hence denoted here $h_{\min}$: 
$\chi(q)=G(h_{\min})$.
Since this value of $h$ is the same for all $q$'s, the result  is a linear behaviour:
\begin{equation}
                \chi(q) = \frac{q+c(h_{\min})}{h_{\min}} 
      \label{eq34}
\end{equation}
The singularity $h_{\min}$ is associated to small moments and the smallest fluctuations 
detectable in experimental time series; it is not clear whether this maximum singularity
reaches a fixed value for Eulerian turbulence, and what is this value. This maximal
singularity also depends on the precision of the experimental measurements which are
analyzed. To check this, we consider numerical and experimental analysis of such inverse statistics for
$q>0$. We plot in Fig. 5 the data
for $\chi(q)$ with $q>0$ reported in Refs. \cite{Jensen1999,Beaulac2004}. The latter are 
experimental estimates, while the former are obtained from shell model calculations.
Both are rather linear (the published experimental estimates contain only the first
three moments), and seem to confirm a linear behaviour such as Eq.(\ref{eq34}).
Recently inverse statistics have been tested using wind-tunnel experimental 
data, confirming a linear behaviour for positive (relative) exponents \cite{Pearson2005}.  

Let us now focus on negative moments of the time statistics: $q>0$ in Eq.(\ref{eq31}). 
This corresponds to positive
moments of inverse times. We have $\zeta(q_0)=-q$  and $\chi(q)=G(h_0)=-q_0$. 
The result will be simply written introducing $p=-q$ and $q_0(p)=-\chi(q)$:
\begin{equation}
        < \left( \frac{1}{\tau} \right)^p> \sim \delta^{-q_0(p)}
     \label{eq35} 
\end{equation}
with:
\begin{equation}
       \zeta(q_0) = p
     \label{eq36} 
\end{equation}
The exponent obtained with the inverse statistics is in fact the inverse
 (in the sense of reciprocal) of the structure function:
\begin{equation}
       q_0(p)  = \zeta^{(-1)}(p)
     \label{eq37} 
\end{equation}
For Eulerian velocity (with Taylor's hypothesis to interprete time statistics as being close
to spatial statistics), a time series measured at a fixed location will be
characterized by $\zeta = \zeta_E$.
Thus $q=1$ selects $q_0 = 3$, non affected by intermittency corrections. The non-intermittent
curve is $q_0=3q$, and taking into account intermittency we have also the
approximate value $q_0(2) \simeq 7$.
The corresponding curves for Eulerian turbulence and for passive scalar turbulence
(data from Ref. \cite{Schmitt2005b} corresponding to an average of several experimental estimates) 
are shown in Fig. 6: they are increasing fast, especially for the passive scalar  case.
Let us note also that the experimental results published in \cite{Pearson2005}
do not confirm Eq.(\ref{eq37}), except for $p=1$. 

Exit-time statistics have also been used in Ref. \cite{Abel2000a,Abel2000b,Boffetta2002b}
to characterize the $\epsilon$-entropy of turbulent signals. The concept of $\epsilon$-entropy
has been introduced
to quantify the predictability and degree of complexity of continuously varying signals such
as turbulent signals. The continuous signal is transformed into a symbolic time series using
a   grid in the phase space, of resolution $\epsilon$. The classical Shannon entropy $h$ of the
resulting symbolic time series is then studied, as a function of the grid size $\epsilon$ (see
\cite{Wang1992,Gaspard1993}).  
Abel et al. \cite{Abel2000a,Abel2000b} have proposed to study the $\epsilon$-entropy of
turbulent signals $h(\epsilon)$ using the exit-times $t(\epsilon)$, the time for the signal to
undergo a fluctuation of size $\epsilon$.
They obtained the following 
expresion
\begin{equation}
       h(\epsilon)  = h^{\Omega}(\epsilon) <\frac{1}{\tau}>
     \label{eq38} 
\end{equation}
where $h^{\Omega}(\epsilon)$ is the exit-time $\epsilon$-entropy, being bounded and having only a logarithmic 
variation with $\epsilon$. The leading $\epsilon$ behaviour of $h(\epsilon)$ is then
given by $<\frac{1}{\tau}>$, where the average is done here by considering a succession of
exit time events, and is thus related to Eq. (\ref{eq35}). This gives using Eq.(\ref{eq35}) and 
$\epsilon = \delta$:
\begin{equation}
       h(\epsilon)  \sim  \epsilon^{-\beta}
     \label{eq39} 
\end{equation}
with $\beta = q_0(1)$ as given by Ref. \cite{Abel2000a,Abel2000b}. But we can write also
using Eq.(\ref{eq37}):
\begin{equation}
       \beta  = \zeta^{(-1)}(1)
     \label{eq40} 
\end{equation}
Hence for a turbulent time series recorded at a fixed location, but using Taylor's hypothesis,
corresponding to Eulerian turbulence, $\beta=3$ \cite{Abel2000a,Abel2000b,Boffetta2002b}.
On the other hand, for Eulerian temperature turbulence (still with Taylor's hypothesis), 
one will have a larger value since $\zeta_{\theta}(3) \simeq 0.8$; using the values
published in Ref. \cite{Schmitt2005b}, we have $\zeta_{\theta}(4) \simeq 1$
and hence we obtain for passive scalar Eulerian turbulence $\beta_{\theta} \simeq 4$.
For times series of Lagrangian velocity and passive scalar turbulence, 
both characterized by $\zeta(2)=1$, we have $\beta=2$. This recovers the result 
obtained in Ref. \cite{Wang1992}, which have been obtained using a dimensional argument,
thus corresponding implicitely to Lagrangian velocity.
Let us also note that Eq.(\ref{eq40}) recovers the results of Kolmogorov \cite{Kolmogorov1956}:
for Brownian processes with scaling power spectra of the form $E(\omega) \sim \omega^{-(1+2H)}$,
he gave $\beta=1/H$. Equation (\ref{eq40}) gives the same result, since for such process,
$\zeta(q)=qH$, hence $\beta  = \zeta^{(-1)}(1)=1/H$.

\section*{Conclusion}
We have introduced a simple procedure helping to express explicitely scaling exponents
expressed as Legendre transforms in multifractal turbulence. We have applied this 
idea to several problems involving intermittent corrections to predictability
and dispersion studies.  Even if the basic idea is simple, this produced several interesting 
new results: several functions are obtained
as parametric transform of the Eulerian velocity structure function. We have confirmed and
generalized some previously known results and provided some predictions.

More precisely, in this paper we have expressed the scaling moment function of infinitesimal
perturbations, that characterize the inverse times statistics in the intermediate
dissipation range, as function of the Reynolds number.
For finite size perturbations, we have also provided the scaling moment function $\beta(q)$ of
inverse times, as function of the size $\delta$ of perturbations, showing that this exponent
is very simply related to the reciprocal of the velocity structure functions exponent
$\zeta(q)$. We veryfied the known result $\beta(1)=2$ and obtain the approximate value $\beta(2) \simeq 5$.
We have also considered Lagrangian velocity structure functions $\zeta_L(q)$ and showed
how to relate them explicitely to the Eulerian curve $\zeta_E(q)$. We compared the
predicted Lagrangian curve to recent experimental values. We noticed the approximate value
$\zeta_L(5) \simeq 1$. We also considered the scaling exponent $H(q)$ for pair dispersion
which is very simply related to the Lagrangian scaling exponents. The obtained curve was
compared to recent numerical estimates. We finally considered exit-time moments, and 
confirmed in the general case a result obtained recently on experimental grounds
and analytically for a special case. We showed that,
for exit-time positive moments, the scale invariant function $\chi(q)$ may often be linear,
due to finite precision of experimental measurements. This is in agreement with several
published experimental results. We then proposed to focus on
inverse statistics considering negative moments, hence positive moments of inverse times.
In this framework we showed that the resulting exponent $q_0(q)$ is the reciprocal of
$\zeta(q)$. Using some recent results linking exist time
statistics to $\epsilon$-entropy expression, we obtain a new general result for the
$\epsilon$-entropy exponent of multi-affine signals $\beta = \zeta^{(-1)}(1)$ 
confirming $\beta=3$ for Eulerian turbulence, and giving as 
prediction $\beta=2$ for Lagrangian turbulence (velocity and passive scalar), $\beta \simeq 2$ for
financial time series,
and the approximate value $\beta \simeq 4$ for passive
scalar Eulerian turbulence.

\section*{Acknowledgements}
 Useful suggestions by the referees are acknowledged.

\begin{table}
\caption{\label{tab1}Some recent experimental estimations for $\zeta_E(q)$
and the average value used here. Ref. \cite{Benzi1993}: $Re$ from 6,000 to 47,000; 
Ref. \cite{Arneodo1996}: various experiments, $R_{\lambda}$ going from 35 to 5,000;
Ref.  \cite{Antonia1997}: turbulent wake, $R_{\lambda}$=230; 
Ref. \cite{VdW1999}: various experiments, $R_{\lambda}$ going from 340 to 800.}
\begin{tabular}{cccccc}
 q &  \cite{Benzi1993}&  \cite{Arneodo1996}
    &  \cite{Antonia1997} & \cite{VdW1999}
     & Average value \\
\hline
1    &           &          &           &   0.368   & 0.37  \\
2    & 0.71  & 0.70 & 0.69  &   0.694   & 0.70   \\
3    & 1       & 1       & 1       &  1            & 1          \\
4    & 1.28  & 1.25 & 1.29 & 1.282     & 1.28  \\
5   &  1.53 & 1.50 & 1.55 & 1.541      & 1.53  \\
6   & 1.78  & 1.75 & 1.79  & 1.782     & 1.78   \\
7   & 2.01  & 1.97 & 2.01 &  2.007  &  2.00    \\
8   & 2.22  & 2.10 & 2.22  & 2.217  & 2.19   \\
\end{tabular}
\end{table}

\begin{figure}
  \includegraphics[width=\textwidth]{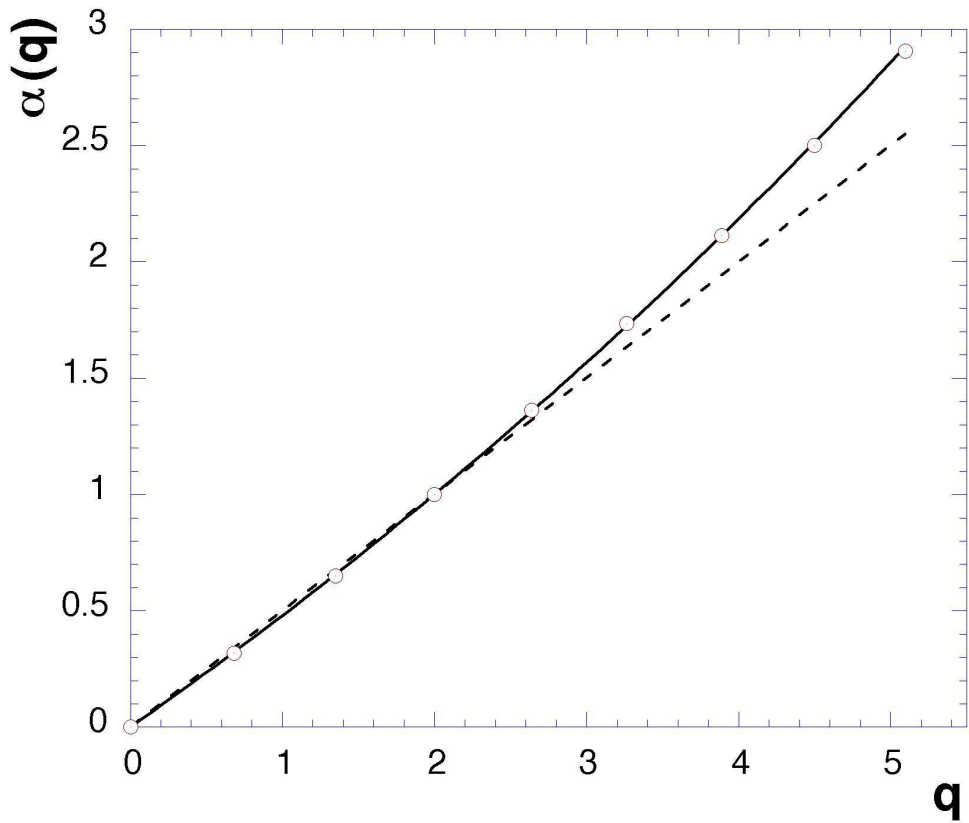}
\caption{\label{fig1} The curve $\alpha(q)$ (open dots: experimental values;
continuous line: lognormal fit) compared to the non-intermittent curve of equation
$q/2$ (dotted line). It is visible that $\alpha(1)=\alpha <1/2$, 
but the intermittency correction is very small.
The value $\alpha(2)=1$ is a fixed point not affected by intermittency. One obtains numerically
$\alpha(3) \simeq 1.56$ and $\alpha(4) \simeq 2.2$.}
\end{figure}

\begin{figure}
  \includegraphics[width=\textwidth]{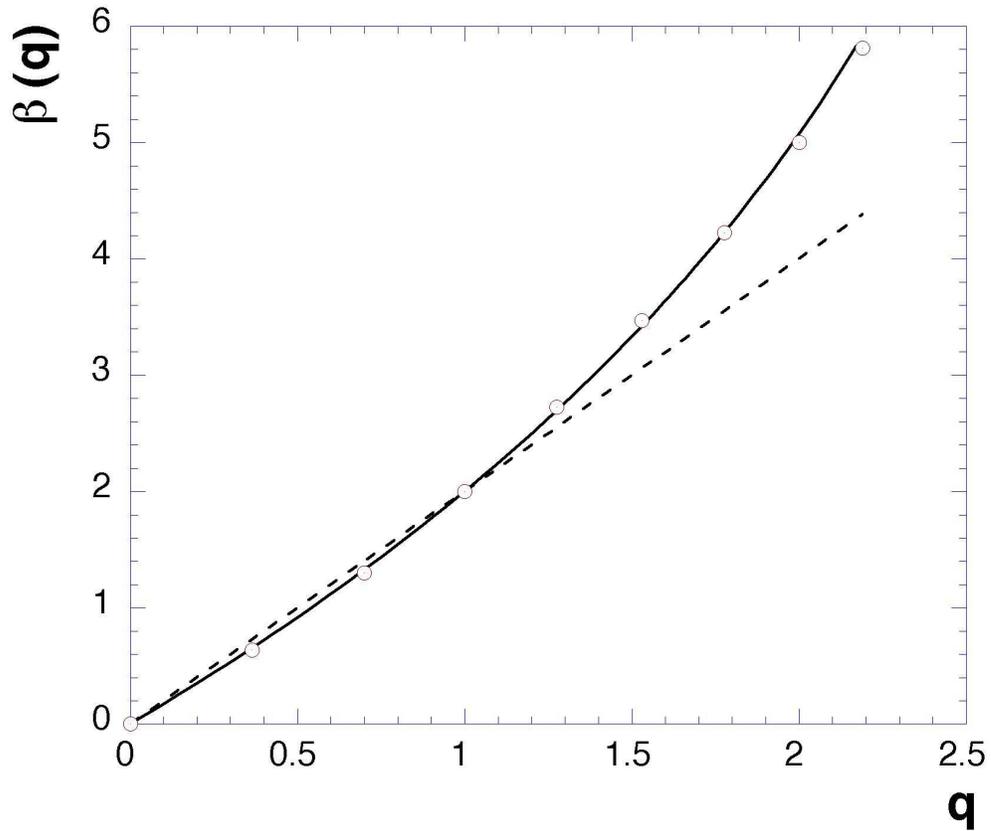}
\caption{\label{fig2} The curve $\beta(q)$ (open dots: experimental values;
continuous line: lognormal fit) compared to the non-intermittent curve of equation
$2q$ (dotted line). This curve is strongly nonlinear and increasing
very fast. The value $\beta(1)=2$ is a fixed point and we may note also that $\beta(2) \simeq 5$.}
\end{figure}

 \begin{figure}
  \includegraphics[width=\textwidth]{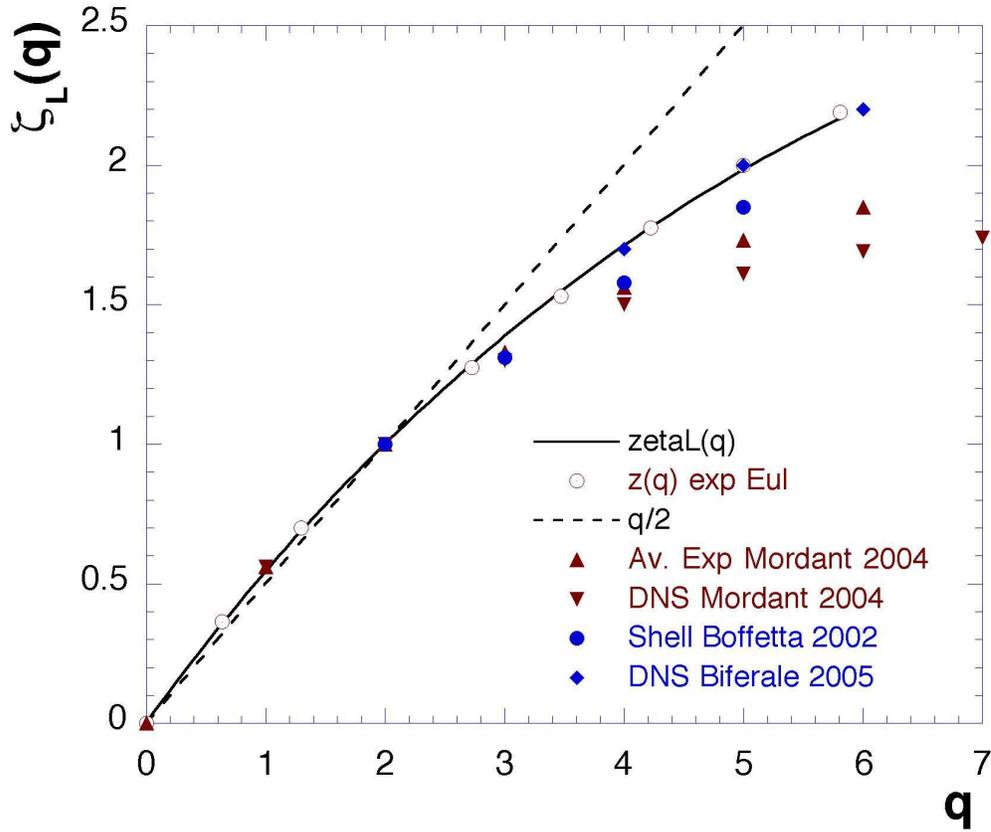}
\caption{\label{fig3} The curve $\zeta_L(q)$ (open dots: experimental values;
continuous line: lognormal fit) compared to the non-intermittent curve of equation
$q/2$ (dotted line), and to four Lagrangian recent 
estimates: average of 4 experiments published in \cite{Mordant2004}; DNS from \cite{Mordant2004}, with $R_{\lambda}$=140; shell model results from \cite{Boffetta2002a}; and DNS from 
\cite{Biferale2005}. The agreement is good for low orders of moment; for larger moments experimental 
and numerical estimates of scaling exponents have a larger scatter, and it is difficult to conclude 
concerning the validity of Eq. (\ref{eq25}).
}
\end{figure}

 \begin{figure}
  \includegraphics[width=\textwidth]{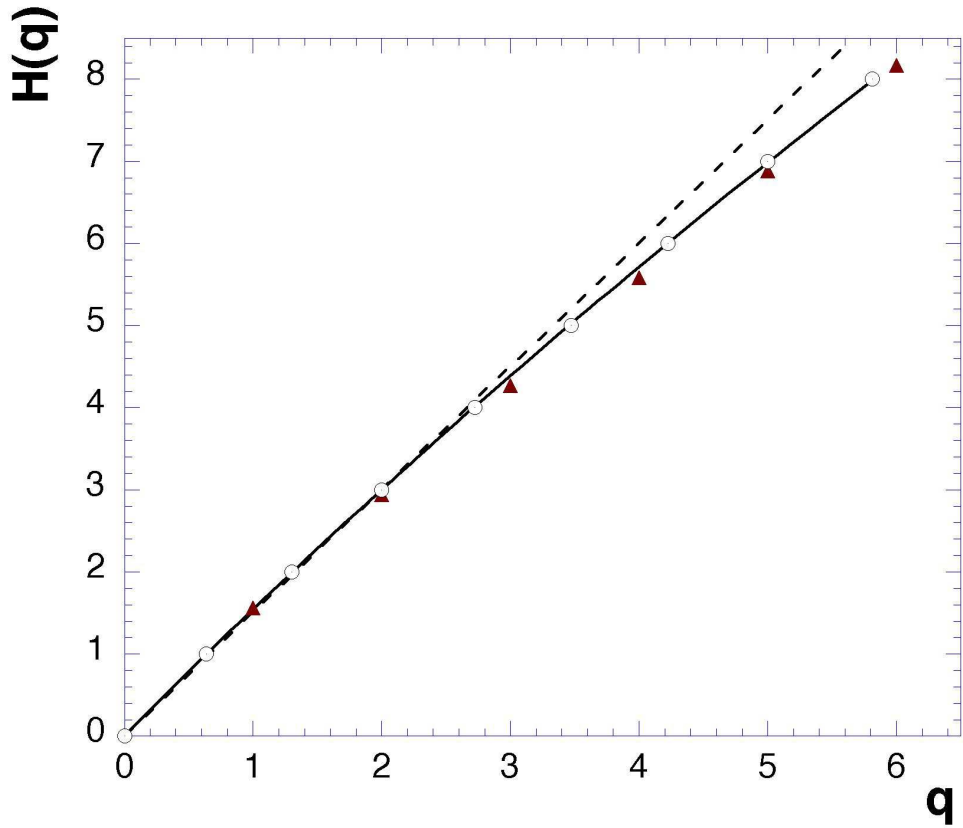}
\caption{\label{fig4} The curve $H(q)$ (open dots: experimental values;
continuous line: lognormal fit) compared to the non-intermittent curve of equation
$3q/2$ (dotted line). This is also compared to the shell-model estimates published in 
\cite{Boffetta1999b}.
The agreement is quite good. The curve $H$ is only slightly nonlinear and concave.
}
\end{figure}

 \begin{figure}
  \includegraphics[width=\textwidth]{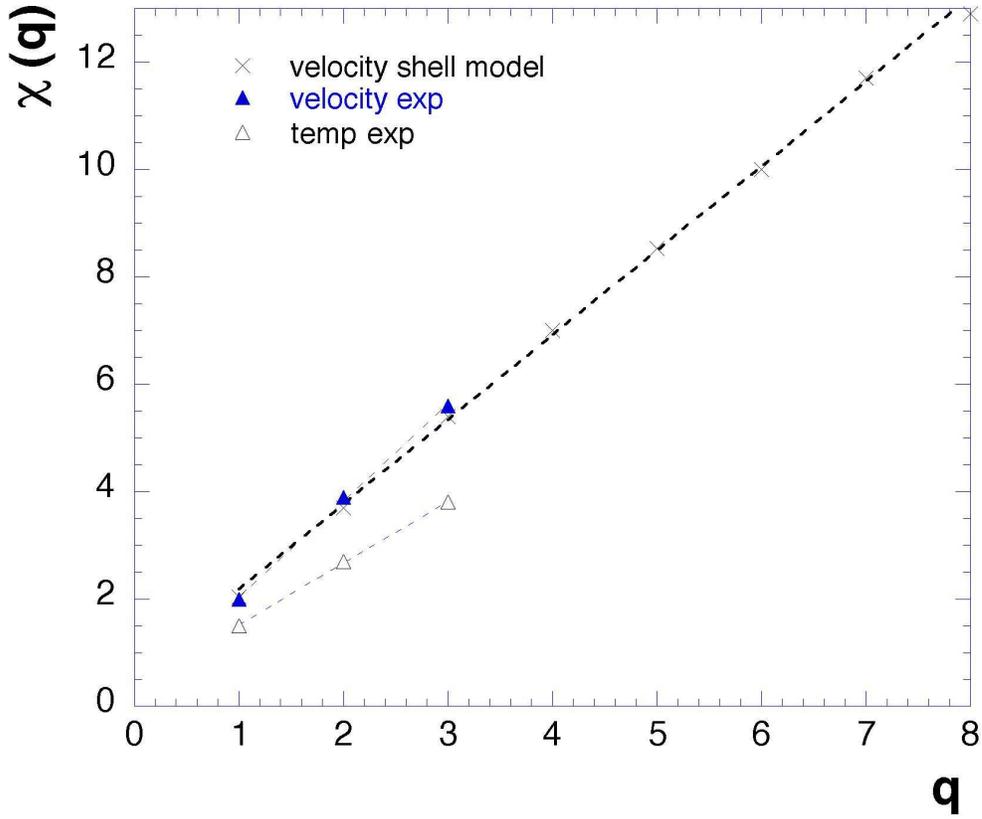}
\caption{\label{fig5} Representation of data for $\chi(q)$ for $q>0$ reported in
Refs. \cite{Jensen1999,Beaulac2004}.
These correspond to numerical estimates from a shell model \cite{Jensen1999} and
experimental estimates with $R_{\lambda}=582$ for velocity (longitudinal structure functions) 
and temperature turbulence \cite{Beaulac2004}. These values are compatible with
a linear behaviour (linear fits are displayed as dotted lines), as 
predicted by Eq.(\ref{eq34}), although experimental estimates
are provided here for only three values.
}
\end{figure}

 \begin{figure}
  \includegraphics[width=\textwidth]{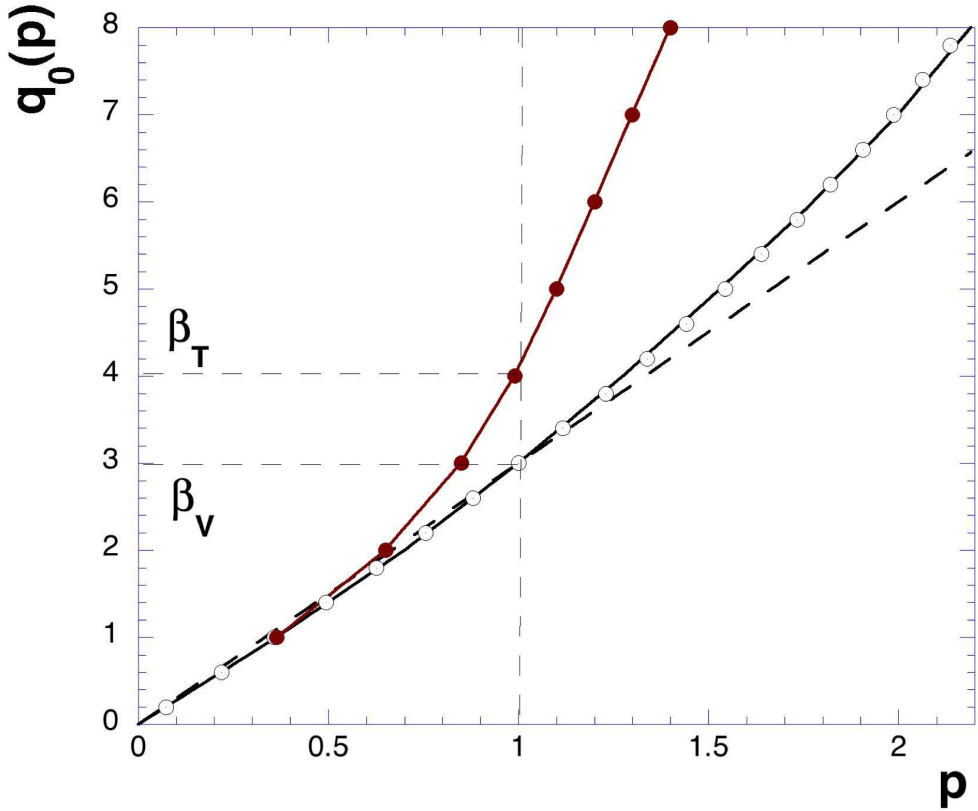}
\caption{\label{fig6} The curve $ q_0(p)  = \zeta^{(-1)}(p)$ representing the scaling exponent
function for the moments of order $p>0$ of the inverse exit times, for the velocity 
(open dots: experimental values;
continuous line: lognormal fit) compared to the non-intermittent curve of equation
$3q$ (dotted line); also shown: the same curve for passive scalar turbulence
(continuous line and closed dots, data from Ref. \cite{Schmitt2005b}). 
The passive scalar curve is very close to the velocity one for weak moments; it increases 
very quickly. We also represent the $\epsilon$-entropy exponent $\beta=\zeta^{(-1)}(1)$,
to emphasize the clear different values for velocity and passive scalar Eulerian turbulence
}
\end{figure}

\end{document}